\documentclass[10pt]{iopart}
\usepackage{iopams} 
\usepackage{graphicx}
\usepackage{epsfig}
\usepackage{float}
\usepackage{amssymb}

\begin{document}
\title[Neutrino hierarchy]{Can we measure the neutrino mass hierarchy
  \\ in the sky?}
\author{Raul Jimenez$^{1,2}$, Thomas Kitching$^{3}$, Carlos Pe\~na-Garay$^{4}$ and Licia Verde$^{1,2}$\\
$^1$ {\it ICREA \& ICCUB-IEEC, University of Barcelona, Barcelona 08028, Spain.} \\
$^2$ {\it Institute for the Physics and Mathematics of the Universe (IPMU), the University of Tokyo, Kashiwa, Chiba, 277-8568, Japan.} \\
$^3$ {\it Institute for Astronomy, University of Edinburgh, Blackford Hill, Edinburgh EH9-3HJ, UK.}\\
$^4$ {\it Instituto de F\'isica Corpuscular (CSIC-UVEG), Val\'encia, Spain.} \\}

\begin{abstract}
  Cosmological probes are steadily reducing the total neutrino mass window,  
  resulting in constraints on the neutrino-mass 
  degeneracy as the most significant outcome.
  In this work we explore the discovery potential of cosmological 
  probes to constrain the neutrino hierarchy, and point out some
  subtleties that could yield spurious claims of detection.
  This has an important implication for next generation of double beta decay experiments,
  that will be able to achieve a positive signal in the case of degenerate or 
  inverted hierarchy of   Majorana neutrinos. We find that cosmological 
  experiments that nearly cover the whole sky could in principle
  distinguish the neutrino hierarchy by yielding `substantial' 
  evidence for one scenario over the another, via precise measurements 
  of the shape of the matter power spectrum from large scale structure and weak  gravitational lensing.
\end{abstract}


\section{Introduction}

In the past decade, there has been great progress in neutrino
physics.  It has been shown that 
atmospheric neutrinos exhibit a large up-down asymmetry in the SuperKamiokande (SK) 
experiment. This was the first significant evidence for a finite neutrino mass \cite{SuperK} 
and hence the incompleteness of the Standard Model of particle physics.
Accelerator experiments \cite{K2K, MINOS} have confirmed this evidence and improved the 
determination of the neutrino mass splitting required to explain the observations.
The Sudbury Neutrino Observatory (SNO) experiment 
has shown that the solar neutrinos change their flavors from the electron type to other active 
types (muon and tau neutrinos)\cite{SNO}. Finally, the KamLAND reactor anti-neutrino oscillation
experiments reported a significant deficit in reactor anti-neutrino flux over approximately 180~km 
of propagation \cite{KamLAND}. Combining results from the pioneering
Homestake experiment \cite{Homestake} and
Gallium-based experiments \cite{Gallium}, the decades-long solar
neutrino problem \cite{solarproblem} has been solved by   the electron
neutrinos produced at Sun's core propagating adiabatically to a heavier
mass eigenstate due to the matter effect \cite{MSW}.

As a summary, two hierarchical neutrino mass splittings and two large mixing angles have been 
measured, while only a bound on a third mixing angle has been established.
Furthermore the standard model has three neutrinos and the motivation 
for considering deviations from the standard model in the form of 
extra neutrino species has now  disappeared \cite{mena,miniboone}.

New neutrino experiments aim to determine the remaining parameters of the neutrino mass 
matrix and the nature of the neutrino mass. Meanwhile, relic neutrinos  produced in the early universe are 
hardly detectable by weak interactions but new cosmological probes offer the opportunity to detect 
relic neutrinos and determine neutrino mass parameters.

It is very relevant that the maximal mixing of the solar mixing angle is excluded at a 
high significance. The exclusion of the maximal mixing by SNO \cite{SNO} has an important impact on a deep  question in neutrino physics: ``are neutrinos their own
anti-particle?".  If the answer is yes, then neutrinos are Majorana fermions; if not,
they are Dirac. If neutrinos and anti-neutrinos are identical, there could have been a
process in the Early Universe that affected the balance between particles
and anti-particles, leading to the matter anti-matter asymmetry we
need to exist \cite{leptogenesis}.  This question can, in principle, be resolved if  neutrinoless double
beta decay is observed.  Because such a phenomenon will violate the
lepton number by two units, it cannot be caused if the neutrino is
different from the anti-neutrino (see \cite{murayama} and references therein).  
Many experimental proposals exist that will increase the sensitivity to such a
phenomenon dramatically over the next ten years (e.g., \cite{0nbb} and references therein).  

The crucial question we want to address is if a negative result from such experiments can
lead to a definitive statement about the nature of neutrinos.  
Within three generations of neutrinos, and given all neutrino
oscillation data, there are three possible mass spectra: a) degenerate, with mass 
splitting smaller than the neutrino masses, and two non-degenerate cases, b) normal hierarchy, 
with the larger mass splitting between the two more massive neutrinos and c) inverted hierarchy, 
with the smaller spitting between the two higher mass neutrinos. 
For the inverted hierarchy, a lower bound can be derived on the effective neutrino 
mass \cite{murayama}. The bound for the degenerate spectrum is stronger than for inverted
hierarchy. Unfortunately, for the normal hierarchy, one cannot obtain a similar
rigorous lower limit.  

Neutrino oscillation data have measured the neutrino squared mass
differences, which are hierarchical. Given the smallness of neutrino masses and 
the hierarchy in mass splittings, we can characterize the impact of neutrino masses  on cosmological observables and in particular on the
\begin{center}
\begin{figure*}[!t]
\includegraphics[width=6.8cm]{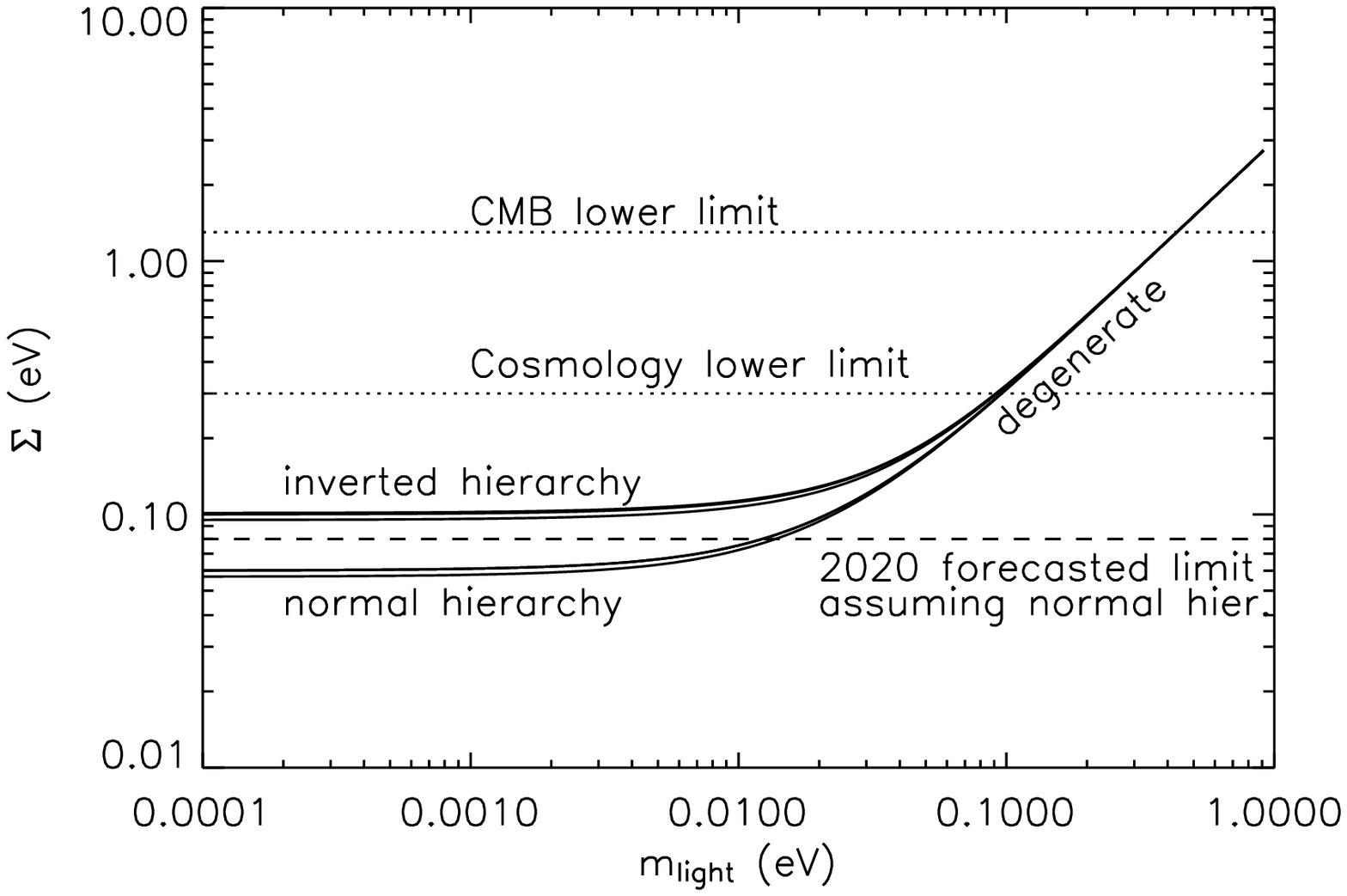}
\includegraphics[width=6.8cm]{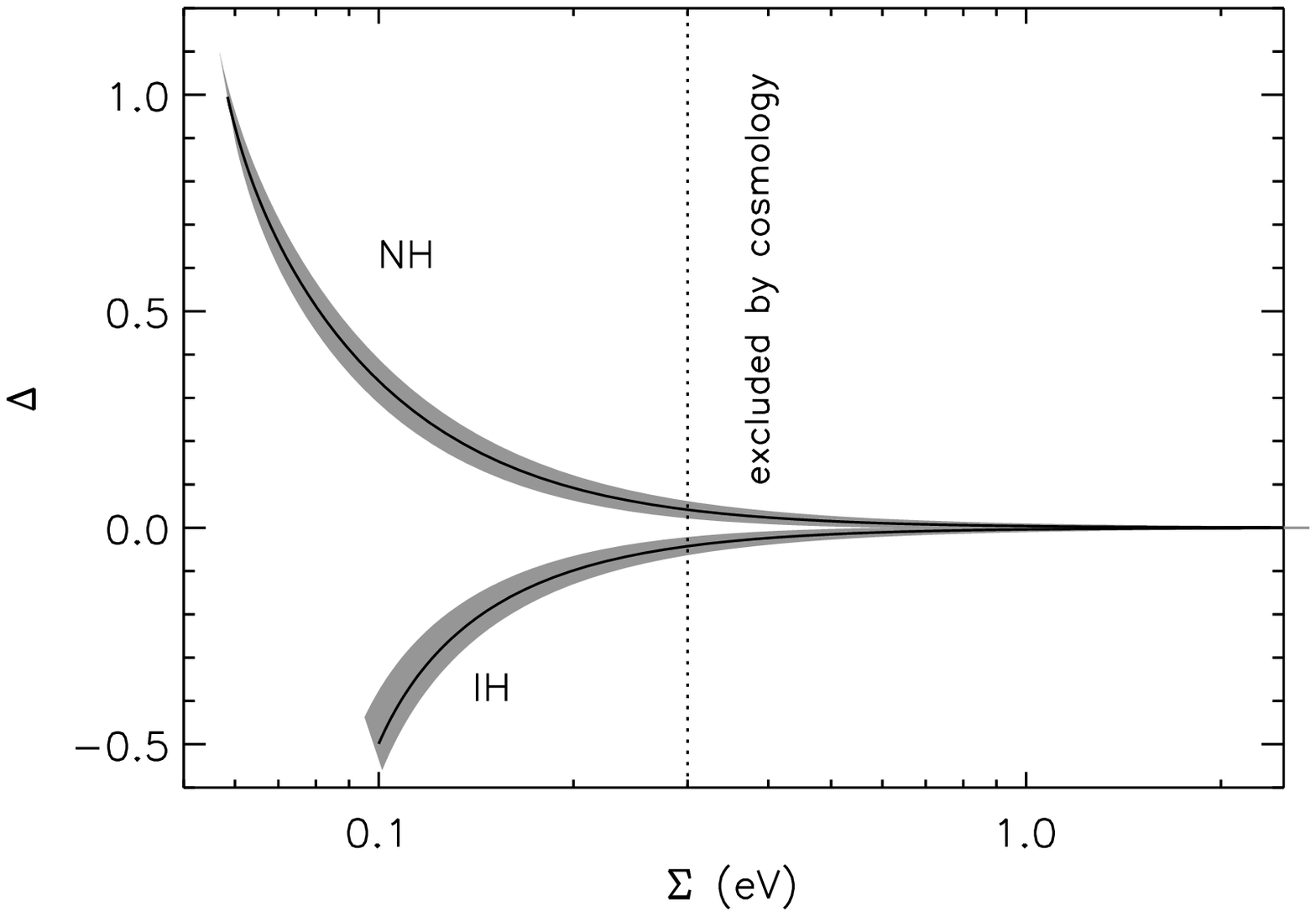}
\caption{Left: constraints from neutrino oscillations and from cosmology in the $m$-$\Sigma$ plane. Right: constraints from neutrino oscillations  (shaded regions) and from cosmology in the $\Sigma$-$\Delta$ plane. In this parameterization the sign of $\Delta$ specifies the hierarchy.} 
\label{fig:0}
\end{figure*}
\end{center}
the matter power spectrum by two parameters: the total mass $\Sigma$ and
the ratio of the largest mass splitting to the total mass, $\Delta$. 
As we will show, one can safely neglect the impact of the solar mass splitting in
cosmology. In this approach, two masses characterize the neutrino mass
spectrum, the lightest one, $m$, and the heaviest one, $M$. 

Neutrino
oscillation data are unable to resolve whether the mass spectrum
consists in two light states with mass $m$ and a heavy one with mass $M$,
named normal hierarchy (NH) or two heavy states with mass $M$ and a
light one with mass $m$, named inverted hierarchy (IH). Near future
neutrino oscillation data may resolve the neutrino mass hierarchy if
one of the still unknown parameters that relates flavor with mass
states is not too small. On the contrary, if that mixing angle is too
small, oscillation data may be unable to solve this issue. 
Analogously, a total neutrino mass determination from cosmology 
will be able to determine the hierarchy only if the underlying 
model is normal hierarchy and $\Sigma<0.1$ eV (see e.g.,  Fig~\ref{fig:0}).
If neutrinos exist in either an inverted hierarchy or are denegerate, 
(and if the neutrinoless double beta decay signal is not seen within
the bounds determined by neutrino oscillation data), 
then the three light neutrino mass eigenstates (only) will be found 
to be Dirac particles.

In this paper, we investigate whether cosmological data may positively
establish the degenerate spectrum from the inverted hierarchy (or vice
versa). Our approach is to take cosmic variance limited surveys,
rather than specifically planned experiments, so that we can determine
if (even in the ideal case) cosmology can make any impact on this
question. 
 
\section{Massive Neutrinos and the Power Spectrum}
\label{Massive Neutrinos and the Power Spectrum}
Massive neutrinos affect  cosmological observations in a variety of different ways.
For example, cosmic microwave background (CMB) data alone constrain the total
neutrino 
mass $\Sigma<1.3$ eV at the 95\% confidence level \cite{Komatsu10}. 
Neutrinos with mass $\lesssim 1$eV become non-relativistic after the 
epoch of recombination probed by the CMB, thus massive neutrinos alter
matter-radiation 
equality for a fixed $\Omega_m h^2$. After neutrinos become
non-relativistic, 
their free streaming damps the small-scale  power and modifies the
shape of the matter 
power spectrum below the free-streaming length. 
Combining large-scale structure and CMB data, 
at present the sum of masses is constrained to be $\Sigma<0.3$ eV \cite{Reidnu}.
Forthcoming large-scale structure data promise to determine the
small-scale ($0.1 \lesssim k\lesssim 1$ h/Mpc) matter power spectrum
exquisitely well  and to yield errors on $\Sigma$ well below $0.1$ eV
(e.g., \cite{LSST, steen, Euclid}).
Here, we assume the standard $\Lambda$CDM model and explore the changes in the
matter power spectra due to the neutrino properties (mass and
hierarchy).

The effect of neutrino mass on the CMB is related to the
physical density of neutrinos, and therefore the mass difference
between eigenstates can be neglected. However individual neutrino
masses can have an effect on the large-scale shape of the matter power
spectrum. 
In fact, neutrinos of different masses have  different transition
redshifts from 
relativistic to non-relativistic behavior, and their individual masses
and their mass splitting change
the details of the radiation-domination to matter-domination regime. 
As a result the detailed shape of the matter power spectrum on scales
$k\sim 0.01$ $h$/Mpc is affected. 
In principle therefore a precise measurement of the matter 
power spectrum shape can give information on both the sum of the
masses 
and individual masses (and thus the hierarchy), 
even if the second effect is much smaller than the first.

We define 
the relation between the neutrino masses $m$ and $M$ and the parameters $\Sigma$ and $\Delta$ as 
\begin{eqnarray}
{\rm NH:} \, \, \,\,\,\,\,\, & \Sigma =  2m + M \,\,\,\,\,\, & \Delta=(M-m)/\Sigma \\
{\rm IH:} \, \, \,\,\,\,\,\, & \Sigma =  m + 2M \,\,\,\,\,\, & \Delta=(m-M)/\Sigma
\end{eqnarray} 
(recall that $m$ denotes  the lightest neutrino mass and $M$ the heaviest).

 In Fig~\ref{fig:0} we show the current constraints on neutrino mass
 properties in the $m$-$\Delta$ and $\Sigma$-$\Delta$ planes.
 While many different parameterizations have been proposed in the 
 literature to account for neutrino mass splitting in a cosmological
 context \cite{slosar,takada,melchiorri} 
 here we advocate using the $\Delta$ parameterization for the
 following reasons. 
 $\Delta$ changes continuously through normal, degenerate and inverted
 hierarchies; 
 $\Delta$ is positive for NH and negative for IH. Finally, as we will show,
 cosmological data are sensitive to $\Delta$ in an easily understood
 way through the largest mass splitting (i.e., the absolute value of $\Delta$), 
 while the direction of the splitting (the sign of $\Delta$) introduces 
 a sub-dominant correction to the main effect. This parameterisation
 is strictly only applicable for $\Sigma > 0$, but oscillations experiments already set $\Sigma>M\gtrsim 0.05$eV.  

It is important to note that not the entire parameter space in the
$\Sigma$-$\Delta$ 
plane (or of any other parameterization of the hierarchy used in the
literature) is allowed  by particle physics
constraints and should be explored: only the regions
around the normal and inverted 
hierarchies allowed by neutrino oscillation experiments should be 
considered (see Fig~\ref{fig:0}).

To gain a physical intuition on the effect of neutrino properties 
on cosmological observables, such as the shape of the matter power
spectrum, 
it is useful to adopt the following analytical approximation, 
as described in Ref. \cite{takada}.
The matter power spectrum can be written as:
\begin{equation}
\frac{k^3 P(k;z)}{2 \pi^2} = \Delta_R^2 \frac{2 k^2}{5 H_0^2 \Omega_m^2}  D^2_{\nu} (k,z) T^2(k) \left ( \frac{k}{k_0} \right )^{(n_s-1)},
\end{equation}
where $ \Delta_R^2$ is the primordial amplitude of the fluctuations,
$n_s$ is the primordial 
power spectrum spectral slope, $T(k)$ denotes the matter transfer
function and 
$D_{\nu} (k,z)$ is the  scale-dependent linear growth function, which encloses the dependence
of $P(k)$ on non-relativistic neutrino species.

Each of the three neutrinos contributes to the neutrino mass fraction
$f_{\nu,i}$ 
where $i$ runs from $1$ to $3$,
\begin{equation}
f_{\nu,i} = \frac{\Omega_{\nu,i}}{\Omega_m} = 0.05 \left ( \frac{m_{\nu_i}}{0.658 {\rm eV}} \right ) \left ( \frac{0.14}{\Omega_m h^2} \right )
\end{equation}
and has a free-streaming scale $k_{{\rm fs},i}$,
\begin{equation}
  k_{{\rm fs},i} = 0.113  \left ( \frac{m_{\nu_i}}{1 {\rm eV}} \right )^{1/2} \left ( \frac{\Omega_m h^2}{0.14} \frac{5}{1+z}\right)^{1/2}  {\rm Mpc}^{-1}\,.
\end{equation}
Analogously, one can define the corresponding quantities 
for the combined effect of all species, by using $\Sigma$ instead of $m_{\nu_ i}$.
Since we will only distinguish between a light and a heavy eigenstate
we will have e.g., 
$f_{\nu,m}, f_{\nu,\Sigma},  k_{{\rm fs},m},  k_{{\rm fs},\Sigma}$
etc., 
where in the expression for  $f_{\nu,m}$ one should use 
the mass of the eigenstate 
(which is the mass of the individual neutrino, or twice as much 
depending on the hierarchy) while in  $k_{{\rm fs},m}$ one should use
the mass of the 
individual neutrino.

The dependence
of $P(k)$ on non-relativistic neutrino species is  in  $D_{\nu} (k,z)$, given by 
\begin{equation}
D_{\nu_i} (k,z) \propto (1-f_{\nu_i}) D(z)^{1-p_i}
\end{equation} 
where $k \gg k_{{\rm fs},i} (z)$ and $p_i = (5- \sqrt{25 - 24
  f_{\nu_i}})/4$. 
The standard linear growth function $D(z)$ fitting formula is taken
from \cite{HuEise}.  

In summary there are three qualitatively different regimes in $k$-space
that are introduced by the neutrino mass splitting
\begin{eqnarray}
 D_{\nu} (k,z)=& D(k,z)  & \,\,\,\,\,\,\, k < k_{{\rm fs},m} \\
 D_{\nu} (k,z)=& (1\!-\!f_{\nu,m})D(z)^{(1-p_m)} & \,\,\,\,\,k_{{\rm fs},m}\!<\!k\!<k_{{\rm fs},\Sigma}\\
 D_{\nu} (k,z)=& (1-f_{\nu,\Sigma})D(z)^{(1-p_{\Sigma})} &\,\,\,\,\,\,\, k>k_{{\rm fs},\Sigma}\,,
 \end{eqnarray}
where the subscript $m$ refers to the light neutrino eigenstate and $\Sigma$ to the sum of all masses.
 
This description is, however, incomplete: the transitions between the
three 
regimes is done sharply in $k$ while in reality the change is very
smooth. 
In addition, the individual masses change the details of the
matter-radiation 
transition which (keeping all other parameters fixed) adds an
additional effect at scales $k>k_{{\rm fs},\Sigma}$.

In order to explore what constraints can be placed on $\Delta$ and
$\Sigma$ for a given survey set-up we can use a Fisher matrix approach. 
The elements of ${\bf F}$, the  Fisher information matrix \cite{Fisher}, are given by 
\begin{equation}
  F_{\theta \gamma}=-2 \left \langle \frac{\partial^2 \ln L}{\partial \theta \partial \gamma}\right \rangle
\end{equation}
where  $\theta$  and $\gamma$ denote cosmological parameters 
(and the Fisher matrix element's indices) and $L$ denotes the likelihood of the data given the model.
Marginalised errors on a parameter are computed
as $\sigma^2 (\theta) = ({\bf F}^{-1})_{\theta \theta}$ .
We can also
calculate expected Bayesian evidence for cosmological parameters using
the approach of Ref.~\cite{HKV,TK10}. In the case that
we are considering we use the formula from \cite{TK10} for the
expectation value of the evidence, in this case the expected Bayes
factor is simply the log of ratio of the Fisher determinants. 
 
Following Ref.~\cite{Seo} the Fisher matrix for the galaxy power spectrum is
\begin{equation}
F_{\theta \gamma} = \frac{V_s}{8 \pi^2} \int_{-1}^{1} d\mu
\int_{k_{\rm min}}^{k_{\rm max}}  k^2 dk N \frac{\partial \ln P(k,\mu)}{\partial \theta} \frac{\partial \ln P(k,\mu)}{\partial \gamma}  
\label{eq:fisher}
\end{equation} 
with $N=[nP(k,\mu)/(nP(k,\mu)+1)]^2$ and $V_s$ is the volume of the
survey. The integration over the projected angle along the light of
sight
\footnote{As it is customary, $\mu$ denotes the cosine of the angle
  with respect to the line of sight.}
$\mu$ is analytical and
the maximum and minimum  wavenumbers allowed depend
on the survey characteristics with the constraint that $k_{\rm max}$ must be in the
linear regime. 
The derivatives are computed at the fiducial model chosen. 
Throughout this paper we assume a fiducial model given by basic
parameters of the standard LCDM 
cosmology \cite{Komatsu10} and the fiducial values for $\Sigma$ and
$\Delta$ 
are then further specified in each case. 

Despite its limitations, the analytic description of the neutrino
effect described above is extremely
useful when performing an order-of magnitude calculation of an
effect. 
Its corresponding Fisher matrix-approach forecasted errors indicate
that 
while $\Sigma$ can be constrained tightly, nearly-ideal, full-sky,
cosmic-variance dominated surveys 
will be needed to obtain promising errors on $\Delta$. 
However, we find that the analytical approximation above overestimates
the neutrino effects 
on the $P(k)$ and therefore under-estimates the forecasted errors, by
factors of $\sim$ few in the 
regime of interest ($\Sigma<0.3$ eV, $\Delta$ along the NI and IH
lines). 
In what follows we therefore use the publicly available 
{\tt CAMB} code \cite{camb} to compute the matter power spectrum.

\begin{center}
\begin{figure}[!t]
\hspace*{2cm}
\includegraphics[width=10cm]{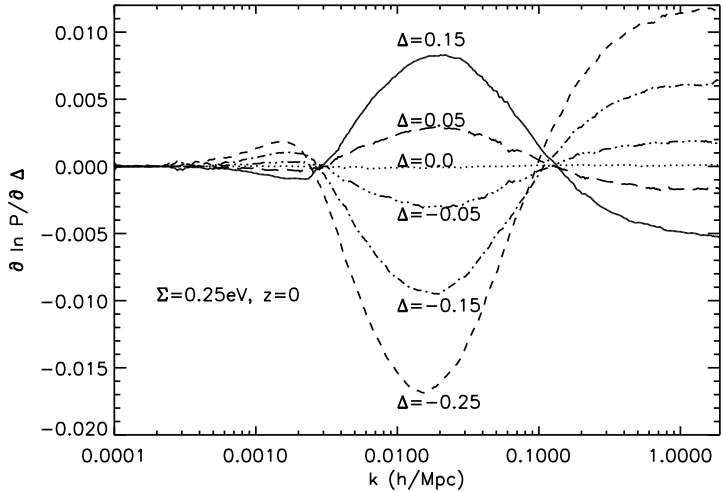}
\caption{Dependence of $P(k)$ on the parameter $\Delta$ at $z=0$, for
  fixed $\Sigma$ and several values of $\Delta$. The dependence is
  expressed as fractional variation in $P(k)$ for a unit variation in
  $\Delta$. For this value of the total mass $\Sigma$, normal
  (inverted) hierarchy correspond to $\Delta \sim 0.05$
  ($\Delta=-0.05$).} 
\label{fig:dpddelta}
\end{figure}
\end{center}

In Fig.~\ref{fig:dpddelta} we show the dependence of $P(k)$ on the
parameter $\Delta$ at $z=0$ for fixed $\Sigma$ and fixed
cosmological parameters. The dependence is shown as the  
 fractional change of the matter power spectrum for a  unit change of
 the parameter $\Delta$. This quantity is then fed directly to the
 Fisher matrix (see Eq. \ref{eq:fisher}). 
  In order  to compute reliably the above derivatives, 
{\tt CAMB} needs to be run at the highest precision settings, 
with fine $k$ sampling and taking care that interpolations 
procedures in-built in the code do not  introduce a spurious signal. 
 
\begin{center}
\begin{figure*}[!ht]
\includegraphics[width=6.8cm,height=7cm]{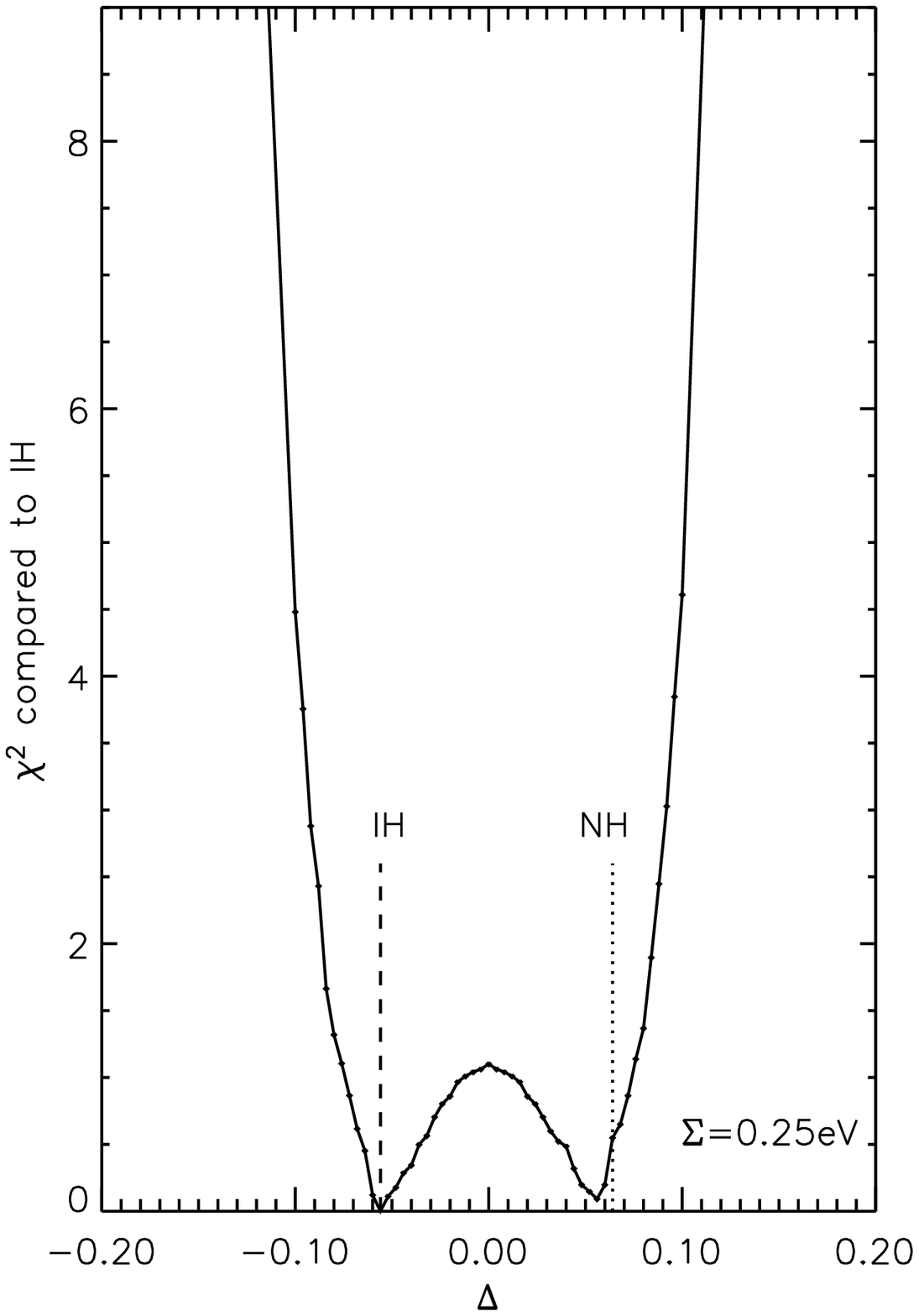}
\includegraphics[width=6.8cm,height=7cm]{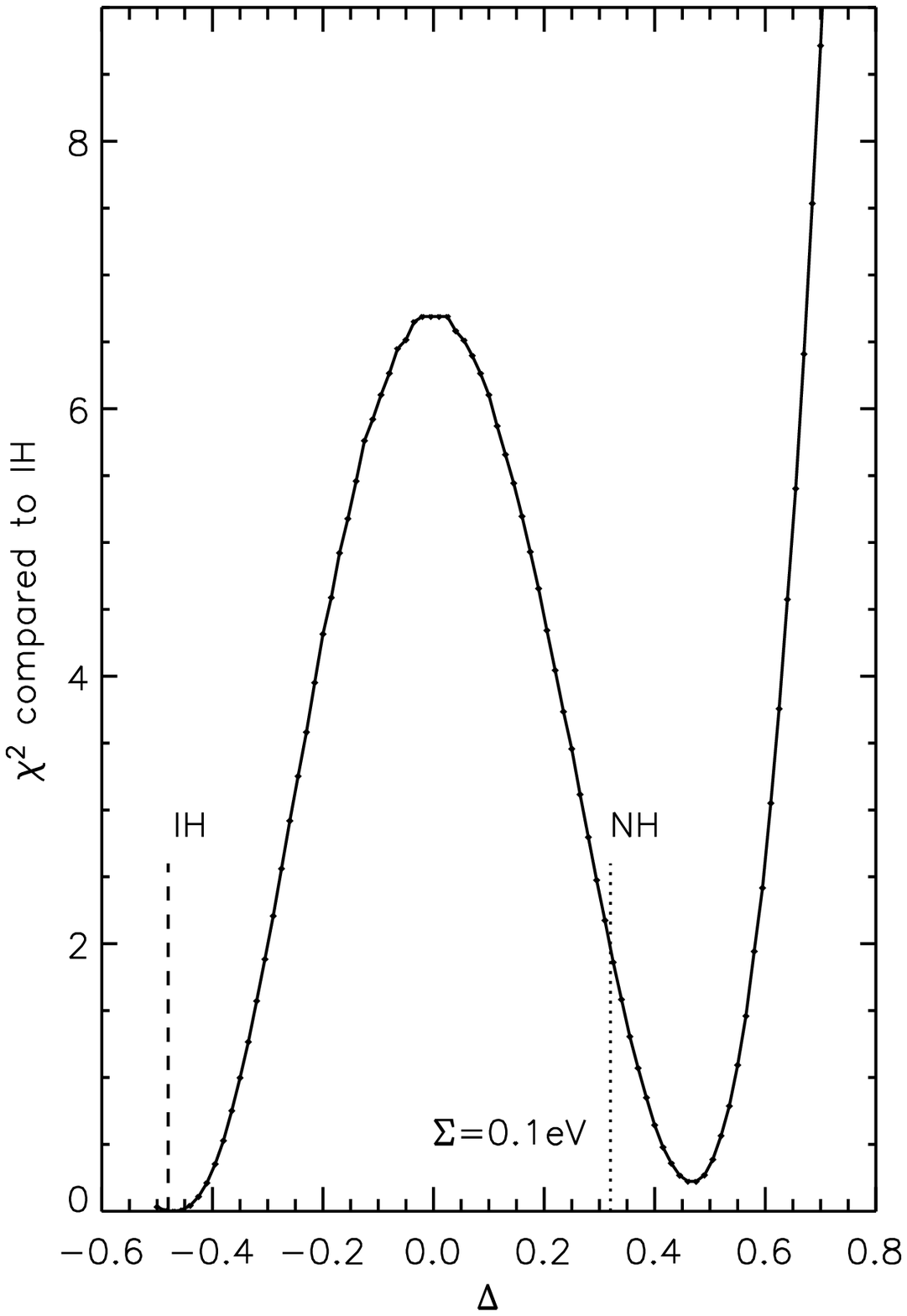}
\caption{$\Delta \chi^2$ as a function  of  the degeneracy parameter
  $\Delta$ for a fixed total neutrino mass $\Sigma$ (and fixed
  cosmology). This is a section  along a  $\Sigma$=constant  line of
  Fig. \ref{fig:0} of the quantity $-2\ln L$ as it would be seen by a
  Fisher matrix approach for a IH fiducial model. The vertical lines
  show the location of the normal and inverted hierarchy. Note the
  bimodal distribution of the $\ln L$ surface, which makes the
  determination of the hierarchy from measurements of the shape of the
  power spectrum extremely challenging. The $\Delta \chi^2$
  normalization matches that achievable from an ideal weak lensing
  survey as described in the text.} 
\label{fig:deltachisq}
\end{figure*}
\end{center}

For the value of the total mass $\Sigma=0.25$ eV, adopted in
Fig.~\ref{fig:dpddelta}, normal (inverted) hierarchy correspond to
$\Delta \sim 0.05$ ($\Delta\sim -0.05$), indicating that the effect of
the neutrino mass splitting on the $P(k)$ is at the $\sim 0.2\%$
level.  
The dependence of $P(k)$ on $\Delta$ at $k>0.1$ $h$/Mpc arises 
because even for a fixed $\Sigma$ the individual masses affect 
the tail of the energy distribution of the relativistic species 
and thus matter-radiation equality.
Note that $\partial \ln P/\partial \Delta$ changes sign with $\Delta$ 
and there is a location, $\Delta=0$, the degenerate case, where $P(k)$
shows no dependence on $\Delta$. 
 
To understand the meaning and implications of this let us 
consider that the 
error on $\Delta$ is directly proportional to $\Delta \chi^2 = -2(\ln
L-\ln L_{\rm fiducial})$ 
where $L_{\rm fiducial}$ denotes the Likelihood for the fiducial model. In addition  we can write
\begin{equation}
  \Delta \chi^2\propto \int_{k_{\rm min}}^{k_{\rm max}}k^2 [P(k,\Delta)-P(k,\Delta_{\rm fiducial})]^2dk\,.
\end{equation}
 
This quantity is shown in  Fig.~\ref{fig:deltachisq}, 
where the normalization has been chosen to match the constraints
achievable from 
an ideal full sky weak lensing survey as the one considered in \S \ref{constraints}.

Fig.~\ref{fig:deltachisq} shows that the likelihood surface is
bimodal:  for example, 
for a fixed cosmology and fixed $\Sigma$, if the fiducial model is the
inverted hierarchy $\Delta <0$, 
there is a corresponding value of $\Delta>0$ (normal hierarchy),
consistent with the neutrino oscillations in the allowed region, with
a  $P(k)$ virtually indistinguishable from the fiducial model. 

The bimodality of the likelihood surface also implies that the
Fisher matrix approach to forecasting errors need to be applied with
care before it can interpreted in terms of distinguishing the
hierarchy. The curvature of the likelihood around the fiducial
model gives the formal error on $\Delta$, and this error may be much
smaller than the distance between IH and NH $\Delta$ values. But this
could be interpreted as a determination of the hierarchy if and only if the
likelihood had a unique maximum, which is not the case here. This subtlety
has not be noticed in the literature before where Fisher
error-estimates for parameters describing neutrino hierarchy were
presented. In general, errors have been computed around one or more
fiducial models and were sometimes found to be smaller than the
distance between normal and inverted. We point out here that this
cannot be directly interpreted as being sufficient to distinguish the
hierarchy (it is a necessary  but not sufficient condition if the
likelihood is multi-peaked). 
 
A more detailed inspection of Fig.~\ref{fig:deltachisq} also indicates
that the $\Delta \chi^2$ between the two minima (maxima of $\ln L$) is
not exactly zero, but it is  very small,  and that the location of the
second minimum (assuming a fiducial IH) does not coincide with the
central value of the oscillations-experiments regions.  
The evidence ratio can the be used to quantify wether a
given survey set up could  distinguish the two cases. 

The philosophy of the rest of the paper is therefore: ``can cosmology in
the cosmic-variance limit, and for an ideal experiment, distinguish the
neutrino heirarchy?" or in other words, ``is there enough 
information in the sky to measure the neutrino hierarchy? "

\begin{center}
\begin{figure*}[!t]
\includegraphics[width=7cm]{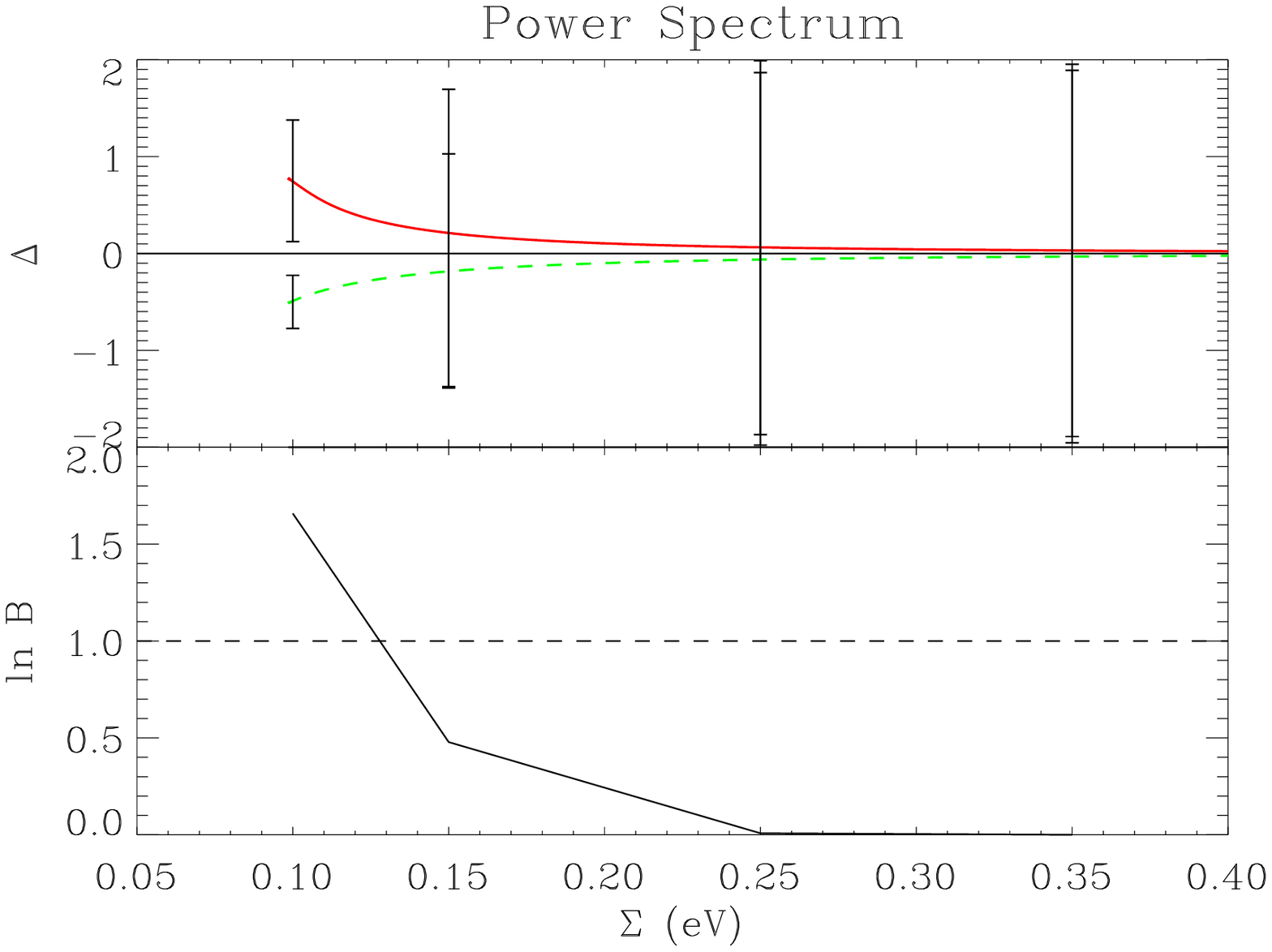}
\includegraphics[width=7cm]{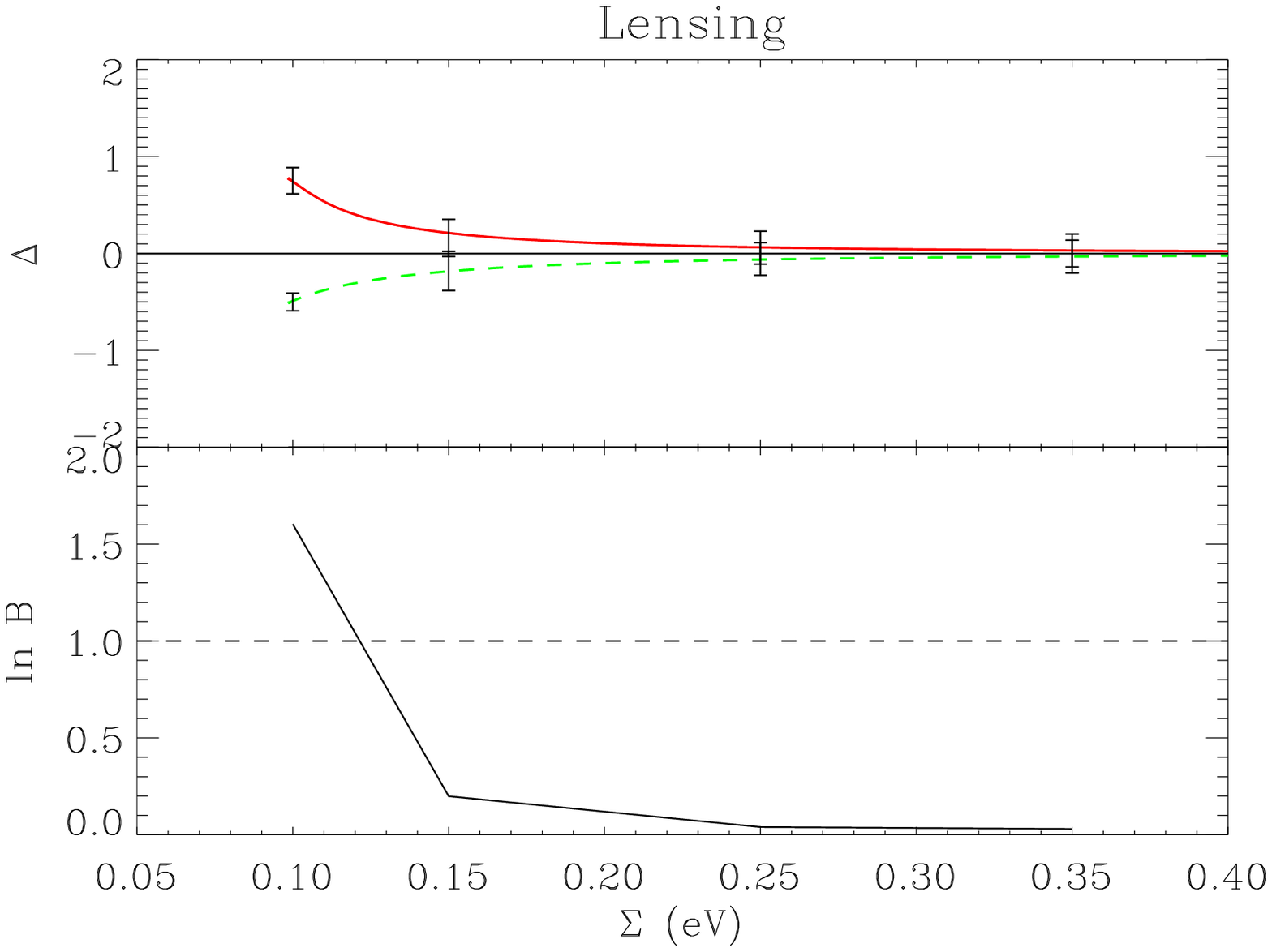}
\caption{LSS (left) and Weak Lensing (right) 
    forecasts for neutrino mass parameters $\Sigma$ and
    $\Delta$. We assume the LSS survey
    consists of a comoving volume of $600$ Gpc$^3$ at $z=2$ and $2000$ Gpc$^3$ at
    $z=5$. The Weak Lensing survey covers $40$,$000$ sq. deg. with a
    median redshift of $3.0$ and a number density of $150$ galaxies per
    square arcminute.
    Several
    fiducial models ($\Sigma$,$\Delta$) were used to derive by Fisher matrix
    approach the expected errors. The upper panel show the $1$-$\sigma$
    errors on $\Delta$ and $\Sigma$, the errors in $\Sigma$ are so small
    that are barely visible. 
    The lower panel shows the expected evidence ratio between the normal and 
    inverted constraints as a function of neutrino mass. The dashed line
    shows the $\ln B=1$ level: in Jeffrey's scale $\ln B<1$  is
    `inconclusive' evidence, 
    and $1<\ln B<2.5$ corresponds to `substantial' evidence.} 
\label{fig:gal}
\end{figure*}
\end{center}

\section{Forecasted Constraints from Large Scale Structure}
\label{constraints}

Here we explore what constraints can be placed on $\Delta$ and
$\Sigma$ from ideal, cosmic variance-dominated future surveys probing 
the shape of the matter power spectrum. 
The two probes of large-scale structure (LSS) we consider are the
matter power spectrum itself and weak lensing.

We also compute the Fisher matrix of a CMB  experiment like 
Planck \footnote{www.sciops.esa.int/PLANCK/} in order to help break
degeneracies in the cosmological parameters when determined only by
the power spectrum, or weak lensing. Therefore our final Fisher matrix is $F =
F_{P(k),WL} + F_{CMB}$. We compute the combined Fisher matrix for variations
in the following cosmological parameters: $n_s, \alpha_s, \Omega_\nu
h^2, \Delta, Z, \Omega_bh^2, \Omega_ch^2, h, A_s$, where $\alpha_s$ denotes the running of the  power spectrum  spectral slope and $Z$ is related to the optical depth to the last scattering surface via $Z=\exp(-2\tau)$; $\Omega_\nu$ is related to $\Sigma$
 via $\Sigma=94 \Omega_\nu$ (eV). 
The reported errors on $\Sigma$ and $\Delta$ are marginalized over the other  cosmological parameters.
The marginal errors for $\Delta$ and $\Sigma$
are shown in Fig.~\ref{fig:gal}; 
the left panel is for a direct $P(k)$ measurement approach and the
right panel is the 
weak lensing approach. 

Because we are interested in answering the question: ``is there enough
information in the sky to measure the neutrino hierarchy?" we have
chosen survey parameters that are ambitous cosmic variance-limited surveys.
For the parameter points shown in the
left panel of Fig. \ref{fig:gal} we have assumed a survey that covers
the full sky $40$,$000$ square degrees 
and maps the positions of galaxies up to $z=2$ corresponding to about
$600$Gpc$^3$ comoving volume and maps
the 21cm-HI up to $z=5$, corresponding to about 2000 Gpc$^3$ comoving
volume.  
We also assumed a high number density of galaxies so that we work in
the cosmic variance-dominated regime ($nP\gg 1$). 
Galaxies are expected to be a {\it biased} tracer of 
the dark matter distribution, here we assume  the bias to be scale and redshift-independent 
and thus not to affect the recovery of the shape of the matter power spectrum. 

HI surveys 
\cite{FurlanettoOhBriggs} target the hyperfine transition in the
hydrogen atom, which in the rest-frame emits a photon in the radio
wavelenghts (21 cm). Therefore they survey the amount of neutral hydrogen in
the universe. Because most galaxies and dark matter overdensities contain
neutral hydrogen, such surveys provide the most un-biased indirect
tracer of the dark matter distribution in the Universe. Further,  in this frequency band, the
radio spectrum is featureless with the only line being the 21 cm
one, its observed frequency yielding a redshift and thus the radial
distance of the emitter. Thus, an imaging survey automatically gives a
three dimensional map of the HI distribution. 
The main challenge facing the HI surveys is the contamination by foregrounds \cite{FurlanettoOhBriggs}.
For the characteristics of the survey we have followed the numbers given in
\cite{Abdalla} which yield to bias of $1$ and negligible shot noise. 

The survey considered is certainly a challenging one, but our calculations indicate that a
cosmic variance-limited galaxy and HI survey can provide enough information
to determine the neutrino hierarchy. We find that such a survey could
constrain the total sum of neutrino mass with extreme accuracy ${\mathcal
  O}(10^{-5})$. We also find that if the total neutrino mass is smaller
than $0.15$ eV, then the IH could be distinguished from the NH through
an evidence criteria centered on each peak in $\Delta$.

Weak lensing is the effect where the path of photons propogating from
a galaxy are distorted by intervening mass concentrations. The amount
of distortion depends on the density and distribution of the
mass. For an individual galaxy image the weak lensing effect is to
induce a change in ellipticity or `shear'. By using redshift and
shear measurements from every galaxy, information on the growth of
structure and the geometry of the Universe can be extracted from 3D
cosmic shear observations. Here we will use the 3D cosmic shear
approach \cite{castro,h03,h06} where the full 3D shear field is 
characterised using 3D spherical harmonics and the 
Fisher matrix methodology of
\cite{h06}.
In line with the cosmic-variance limited approach of this article, 
we assume a large, cosmic-variance limited weak lensing survey
covering $40$,$000$ square degrees, to a median redshift of $3$ with
$50$ galaxies per square arcminute. 
 On the right panel of  Fig. \ref{fig:gal}, we show the marginalised constraints $\Delta$ and
$\Sigma$, for this cosmic-variance limited survey. We find that the
sum of neutrino mass is constrained to extreme accuracy ${\mathcal
  O}(10^{-6})$. As the neutrino mass decreases the constraints on the
IH and NH become smaller and for massess below $\sim 0.15$ eV the
evidence ratio for the IH and NH constraints (lower panels of Fig. \ref{fig:gal}) would become substantial
(in a Jeffrey's scale), allowing for the neutrino heirachy to be
distinguished. This is again a very challenging survey, and acts to
show highlight how demanding the measurement of neutrino
mass-splitting can be; however by using shear measurements from
Euclid \cite{Euclid} or LSST \cite{LSST} we may hope to approach this
regime. 

The degeneracies between $\Sigma$ and $\Delta$ are small, and the very
small constraint on $\Sigma$ results in the constraints being
effectively un-correlated in the $\Sigma$-$\Delta$ plane. 
We note that the constraints on $\Delta$ around the IH and
NH peaks are tighter for weak lensing than LSS, this is due to
lensing providing constraints on both the geometry and the growth of
structure, which provides a smaller raw constraint and a more
orthogonal constraint to the CMB resulting in smaller
errors. Interestingly, even though the weak lensing constraints on
$\Delta$ are smaller than for the power spectrum, the evidence ratio
is comparable, because, 
due to the multi-dimensional degeneracy directions, 
a naive correspondence between error-bars and evidence is not
applicable (it is to a first approximation the difference between the two error
bars that is important).

Note that the evidence
calculation explicitly assumes two isolated peaks, and so is only
applicable when the fiducial points are seperated by multiple-sigma. 
As a result of this, the evidence calculations may be slightly optimisic
for large masses. However, for $\Sigma <0.2$ eV, the $\chi^2$ difference between the two minima becomes noticeable   as well as the shift between the location of one of the two minima  and the central $\Delta$ value for the oscillations experiments (which induces an additional $\chi^2$ difference). While this information  is not fully accounted for in a Bayesian approach to forecasting  the evidence, it may be included at the moment of analyzing the data, using different approaches such as the likelihood ratio, and may slightly improve the significance for  the hierarchy determination.

While we have used the oscillation results to center the Fisher and evidence calculations on the NH and IH, combining the oscillation experiments constraints will not improve the evidence; in fact,  oscillation experiments give symmetric errors  on $\Delta$ (i.e. they do not depend on the sign of $\Delta$).

\begin{center}
\begin{figure*}[!ht]
\includegraphics[width=14cm]{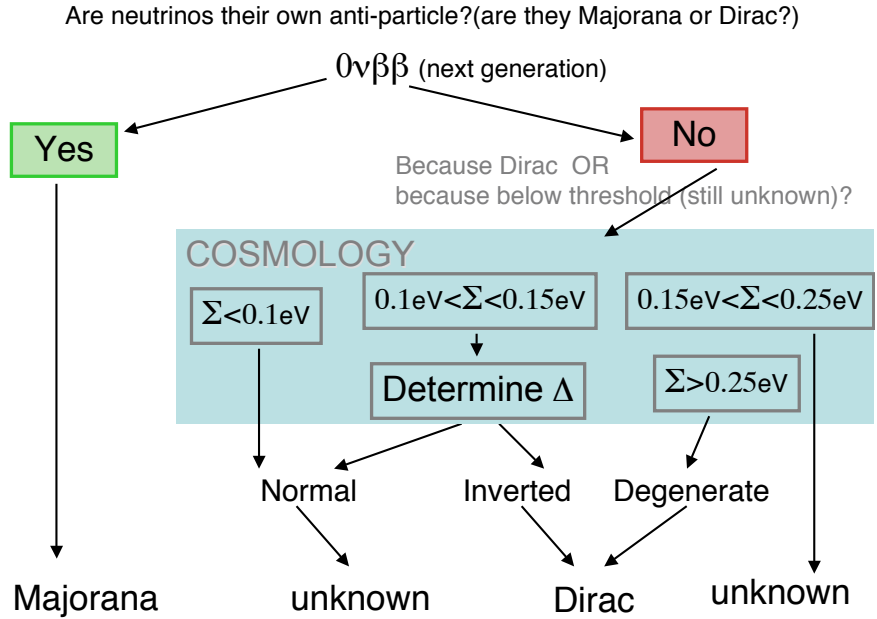}
\caption{Role of cosmology in determining the nature of neutrino mass.  Future  neutrinoless double beta decay ($0\nu \beta beta$)  experiments and future cosmological surveys will be highly complementary in addressing the question of whether neutrinos are Dirac or Majorana particles. Next generation  means  near future experiments whose goal is to reach a sensitivity to the neutrinoless double beta decay effective mass of $0.01$ eV. We can still find two small windows where this combination of experiments will not be able to give a definite answer,  but this region is much reduced by combining  $0\nu \beta beta$ and cosmological observations.} 
\label{fig:flowchart}
\end{figure*}
\end{center}

\section{Conclusions}

The shape of the  matter power spectrum contains information, in
order of decreasing sensitivity, about the sum of neutrino masses,
the amplitude of the mass splitting and the hierarchy (i.e., the mass
splitting order).  
We have introduced a novel parameterization of the neutrino mass
hierarchy, $\Delta$, that has the advantage of changing continuously
between normal, degenerate and inverted hierarchies and whose sign
changes between normal and inverted. The absolute value of $\Delta$
describes the maximum mass difference between the eigenstates. 
We stress that, current constraints from neutrino oscillations have
ruled out large part of the parameter space given by the sum of the
masses and the $\Delta$ parameter, leaving two narrow regions: for a
fixed value of the total mass, the value of  $\Delta$ for the normal
hierarchy is related to that of the inverted one and
$\Delta_{NH}\simeq -\Delta_{IH}$ (but, in detail, $\Delta_{NH} \not \equiv |\Delta_{IH}|$). It is the allowed region that
cosmology should explore. 

We found that the information about $\Delta$ accessible from the power
spectrum shape yields a degeneracy: parameters values $\Delta$ and
$-\Delta$ yield nearly identical power spectra and therefore that the
likelihood surface in $\Delta$ is bimodal.   
This was not noted in the literature before and not taking this
into account when using the Fisher matrix-approach to forecast
future surveys performance may lead to spurious indications of a
surveys ability to determine the hierarchy. 

Detecting the signature of the hierarchy in the sky is therefore
extremely challenging, and therefore we asked: ``can cosmology in the
cosmic-variance limit, and for an ideal experiment, distinguish the
neutrino heirarchy?" or in other words, ``is there enough information
in the sky to measure the neutrino hierarchy?" 
To address these questions we have considered ideal, full-sky, cosmic
variance-limited surveys and found that substantial Bayesian evidence
($\ln B\geq 1$) can be
achieved. 
Are such a surveys feasible in the next $5$-$10$ years? There are two
candidates for such surveys : a full extragalactic survey in the
optical/infrared like Euclid\footnote{http://sci.esa.int/euclid} 
\cite{Euclid} and a
full 21cm survey by the
SKA\footnote{http://www.skatelescope.org}. Each of these surveys is
scheduled to start operations by 2018. Euclid will make an
all sky Hubble-quality map for weak lensing and will directly trace
the dark matter using this technique; whilst the cosmic variance
limited survey we consider here is ambitous with respect to this
survey these result serve as a qualitative measure of this surveys
expected performance (costraints should be only a factor $\leq 1.5$
larger at worst). Euclid will also target
emission line galaxies up to $z \sim 3$ (therefore these galaxies will
have bias parameter close to $1$) 
however $nP$, quantifying the the ratio of the signal to shot noise, will
be only slightly above $1$. The 21cm surveys provide the most
un-biased indirect tracer of the dark matter distribution in the
Universe and have negligible shot noise.

For the degenerate and inverted mass spectra, the next generation
neutrinoless double beta decay experiments  can determine if neutrinos are their own anti-particle.
For the normal hierarchy, the effective electron-neutrino mass may
even vanish. However, if the large-scale structure cosmological data,
improved data on the tritium beta decay, or the long-baseline neutrino
oscillation experiments establish the degenerate or inverted mass
spectrum, the null result from such double-beta decay experiments will
lead to a definitive result pointing to the Dirac nature of the neutrino mass. This is summarized in figure \ref{fig:flowchart}.

If the small mixing in the neutrino mixing matrix is negligible, cosmology might 
be the most promising arena to help in this puzzle. Our work shows that depending 
on the total neutrino mass, there might be substantial evidence by cosmological 
data to infer the neutrino hierarchy.

\section*{Acknowledgments}
CPG is supported by the Spanish MICINN grant
FPA-2007-60323 and the Generalitat Valenciana grant PROMETEO/2009/116.
LV acknowledges support from FP7-PEOPLE-2007-4-3-IRG n. 202182 and FP7- 
IDEAS Phys.LSS 240117. LV and  RJ are supported by MICINN grant AYA2008-03531.  LV \& RJ 
acknowledge support from World Premier International Research Center
Initiative (WPI initiative), MEXT, Japan. TDK is supported by STFC
Rolling Grant RA0888. TDK thanks ICCUB-IEEC, University of
Barcelona and Instituto de F\'isica Corpuscular (CSIC-UVEG),
Val\`encia, for hospitality during part of this work. We thank Alan F. Heavens for
useful discussions.

\section*{References}
\bibliographystyle{JHEP}
\providecommand{\href}[2]{#2}\begingroup\raggedright

\end{document}